\documentclass[12pt]{article}
\makeatletter

\@addtoreset{equation}{section}
\makeatother
\usepackage{amsfonts}
\usepackage{eucal}

\begin{document}

\title{\bf Time representation on the Beginning of space-time \\　\normalsize{{－From a philosophical and mathematical view}}}
\author{Tadashi Fujimoto\\
Department of Philosophy, Ryukoku University\\
Kyoto 600-8268\\
Japan\\
E-mail: fujimoto-t@let.ryukoku.ac.jp}
\date{}
\maketitle

\begin{abstract}
I would like to consider the Beginning of space-time in this paper. First of all, we do consideration historical thought. A lot of philosophers have considered the relation between this real phenomenal world and the basic world in which the phenomenal world is grounded. We will glimpse thought historical details about such respect. Afterwards, we interpret the representation of space-time on Quantum theory and Relativistic theory.  In this case, we will take recent results of time operator theory into consideration. 
\end{abstract}


\section{Introduction}
The Beginning of space-time is the world before the following process. This world is the space of $10^{-40} $cm before our present space-time is born by the ``quantum swing "  of dark energy.  After this swing, ``vacuum" had ``phase transition" and ``inflation" had been caused at $10^{-36} $sec .  And, ``Big Bang" was absorbed as a result. Therefore, we want to consider the world before the birth of the space-time in this paper.  We would not like to argue about what essence our present space-time has. However, at least, according to the common sense in daily life, our living in the 4-dimentional space-time might be incontrovertible to us.  We will not be able to disregard the parameter on time when thinking about the cosmos generation theory \footnote { For example, Big Bang theory. Moreover, we live in the building on the assumption of  one-dimentional time, 3-dimentional space}. 

By the way, we have the language and logical form in the space-time, so how can we talk about the world of non space-time? Can we talk about the non space-time? A lot of philosophers have thought about this world old times. The meaning of the space-time and the origin of the space-time have been considered in natural philosophy and metaphysics. Moreover, mathematics and logic have developed at least as field that doesn't suppose the meaning of the parameter related to the space-time beforehand. 

First, we glimpse how past philosophers have thought about the outside of the space-time and the inside of the space-time. Their thinking processes will give meaningful ideas in considering the logic and the ontology of the Beginning of space- time. 
Next, by invoking mathematics we want to deepen the story in the first half of this paper in order to search for the theory that exceeds theory of relativity and the quantum physics. The physical theory that is existing now might be maybe all powerless to discuss the Beginning of space-time. 

\section{A philosophical viewpoint }
Immanul Kant (1724--1804) had discussed whether cosmos has the start concerning space and time. (cf. A426/B454-A433/B461)\footnote{ Here, A means first version, B is second version. first version 1781, second version 1787. } in the first antinomy theory in \textit{Critique of Pure reason}. According to Kant, this issue can be shown that the following propositions --whether cosmos has the beginning or not, whether cosmos has the start on time or not. As a result, Both propositions don't be appropriate propositions. Kant limited our theoretical knowledge which man is able to share universally and appropriately to the knowledge in our experienced world. Our ability to give empirical and appropriate knowledge are the intuition form of space-time and pure intellectual concepts, according to Kant. The theoretical knowledge is to have a right meaning only in the space-time area, if we can say up to date. Actually, we will doubt the discussion that disregards a space-time form in our scientific theory and daily conversation. In physics, the theory of the non space-time world might fundamentally refuse the means of the experimental verification. 

By the way, how did Kant catch the verge between experience world-phenomenon world (world of space and time) and the super-experience world (non-world of space and time) ? He had pointed out two important points in addition to critique to old thought concerning God's existence proof (cf. A631/B659--A642/B670). 

Briefly to describe Kant's discussion, his insistence is the following.

\medskip

1) ``Existence" in the non space-time world is different from ``Real existence" in the space-time world. And, we cannot enhance the logic on real existence in the space-time to the logic of the existence of the non space-time world. 

\medskip
2) We cannot connect causality that is the effective law in the area of the space-time with non space-time world immediately.

\medskip

The problem about what Kant had talked existed from ancient Greece. 

\bigskip

The cosmology of Plato (BC.427--347) is argued in \textit{Timaeus}. It is said that this writing influenced some natural philosophers of the Renaissance period, especially J. Kepler (1571--1630). Moreover, W. Heisenberg also had touched \textit{Timaeus} in his \textit{Der Teil und das Ganze}(1969). He had described for the regular polygon in \textit{Timaeus} to be compared with the elementary particle. Plato's ontology is based on to the IDEA theory. The IDEA is a basic concept frame in the phenomenal world. The IDEA shows genuine existence at the same time. There was a problem in Plato's IDEA theory,  so Aristotle (BC.384--322) had criticized The IDEA in the paradox of ``The third kind of man" in his \textit{Metaphysics}.  When we limit to the issue of the space-time, the problem is following;  where does connect the IDEA world to the phenomenon world ? 

The phenomenal world is not corresponding to the IDEA world according to Plato. However, Plato thought that things of the phenomenal world is imperfect because they are being put on the space-time world. Plato described that time arises with space and time goes to with the celestial motion (cf. 38b in \textit{Timaeus}). Plato wrote that before time born, a chaos state existed (cf. 30a). Therefore, time comes to generate of our cosmos .

For Plato, the following can be at least said.

\medskip
1) Time comes to exist by the formation of the cosmos and time is related to the movement of the heavenly body.

\medskip
2) Therefore, neither generation (movement) nor time synchronize. So, there is no time in the chaos .

\medskip

 Aristotle caught time in relation to the movement. In other words, time as a representation of the movement (number). Time won't exist in the world where the movement cannot be observed. Moreover, time is existential, because time depends on the movement. J.H.Lambert(1728--1777) who lived in the same age as Kant had succeeded the realism like Aristotle's time concept, contrary to Kant. Though there are a lot of cosmologies of the philosopher to who should pay attention besides these, we omit them this time. 

Should we add any points of time on causality and existence when we talk about the Beginning of space-time?  

\section{Time representation in quantum theory}

We will think about the representation on time of the quantum theory. To avoid falling into useless confusion which is not essential, here we restrict to the Schr\"{o}dinger operator that treats non relativistic quantum system. And, in addition, we limit the model of the movement to the one-dimentional space of one particle as follows. 
Moreover, we restrict the Canonical Commutation Relation (CCR) to the Schr\"{o}dinger representation, as seen as follows. 

CCR is to be filled the following for all ${\psi}{\in} D(QP) {\cap} D(PQ) $, when two self-adjoint operators\footnote{ $D( ･ )$; domain of the operator. Suppose; $D(T) $ is dense. When $T$ were self-adjoint operator, $D(T)=D(T^{*})$ , $T=T^{*} $ hold. It was provided for the first time accurately that the amount of the observable was described with the self-adjoint operator by the von Neumann} 

\hspace{25mm}$[Q,P]\psi=(QP)\psi-(PQ)\psi=i\hbar\psi=i\frac{h}{2\pi}\psi.$ 

\hspace{25mm}$[Q,Q]\psi=0=[P,P]\psi. $

We use $c=1$, ${\hbar}=1$: unit systems. $h$ is Planck's constant. CCR on $Q$,$P$ is $[Q,P]=i$；now ,we consider CCR considering a position operator and a momentum operator. When $Q$，$P$ is Schr\"{o}dinger representation, $Q=M_{x}$, $P=-iD$, $M_{x}$ is multiplication operator on $x$, $D$ is the (generalized, distributional) partial differential operator. When a Hamiltonian  $H$ has potential $V$, the potential operator is represented by $V(Q)$.

Next, now , $\mathcal{H}$ ; Hilbert-space，and we define $\mathcal{H} = L^{2}(\mathbf{R})$. $L^2$ means the whole of square integrable function  (for Lebesgue measure equivalence class).

In this case, Hamiltonian is following;
\begin{equation}
H={\frac{1}{2m}}P^{2}+V(Q),      \hspace{10mm} D(H){\subset}L^{2}(\mathbf{R})
\end{equation}

Observables of quantum systems is is represented by a self-adjoint operator on $ L ^ {2} (\mathbf {R}) $, and it is not necessarily bounded operator. $P$ is the momentum operator (Schr\"{o}dinger representation). $m$ is the mass. $V$ is the external potential. If $V=0 $,  the Hamiltonian $H$ [see (3.1)] is expressed in $ H_ {0} $. The inner product of Hilbert spaces is denoted by $(x , y)$. If real function $V$ were bounded below and Hamiltonian were a smooth infinitely differentiable function,  $H$ become self-adjoint operator on a suitable space. Here we are enough to think about the free Hamiltonian $ H_0 $.

When a self-adjoint Hamiltonian is decided,  for Schr\"{o}dinger type Hamiltonian $ H_0 $ the following equation is determined

\hspace{25mm}${i}\frac{d\psi(t,x)}{dt}=H_0\psi(t,x)$, \hspace{5mm}$\psi(t_0, x)=\psi(x)$

Concerning on state vector $\psi(t,x)$ which $ H_0 $ acts on,  in physical context, $t$ represents time, $x$ represents coordinates (positions).
 $ H_0 $, which plays the role of the state vector generator of time evolution of the system. When $H_0 $ does not contain the time explicitly, it becomes $U (t, t_0) = \mathrm {exp} (- {i} H_0 (t,-t_0)) $;  a unitary operator form.
 The time evolution $\psi (t, x) $ is described as follows. The picture of the evolution of the state is called Schr\"{o}dinger picture.
 
\begin{equation}
\psi(t, x)=U(t, t_0)\psi(t_0, x)
\end{equation}

This represents the continuous change of the state vector on time $t$. And we usually express ${t_0}$ with 0 on $U (t, t_0) = \mathrm{exp} (- {i} H_0 (t,-t_0)) $. The concrete representation of $U (t, t_0) {\psi (x)} = \mathrm{exp} (- {i} H (t, 0)) \psi (x) $ is in general the integral kernel of the propagator representation or the path integral representation. These representation, when the Hamiltonian is not self-adjoint , then meaningless\footnote{propagator or path integral is not intended to give direct quantization. They should be regarded as a specific view of the evolution of the state vector. Strictly speaking, "the existence of infinite-dimensional measure"and "convergence problems of integration" is mathematical almost transcendental method. }.

As state vector representation of Schr\"{o}dinger type evolution, we here express the path integral representation and the propagator integral kernel as follows\footnote {The proof is [10],[12]. }.

For all $\psi{\in}{L^{2}(\mathbf{R})}$ and ${t\in{\mathbf{R}}{\backslash}{0}}$，in means of
${L^{2}(\mathbf{R})}$-convergence，

\begin{equation}
e^{-itH_{0}}{\psi}(x)={\textrm{lim}}_{R{\to}\infty}\biggl(\frac{m}{2{\pi}|t|}\biggr)^{\frac{3}{2}}{\int}_{|y|{\le}{R}}e^{i{\frac{m}{2t}}|x-y|^{2}}\psi(y)dy
\end{equation}

In mathematically strict sense, this is ``Free particle Shr\"{o}dinger-Hamiltonian view" on the integral kernel. We can obtain ``Green function representation"  by using this integral kernel and using the Hilbert space inner product for the resolvent operator.
In the style of physics, we may call propagator as integral kernel[8]. Mathematically, however, we must define a state function which the propagator acts.

Next, For free Hamiltonian path integral representation we apply Trotter formula to this integral kernel representation\footnote{Trotter formula [12];  let $A$, $B$ be self-adjoint operators on a separable Hilbert space. When  $A + B$ is the self-adjoint operator with the domain $D = D (A) {\cap} D (B) $, in the sense of strong convergence $\mathrm {lim} _ {n \to\infty } (e^{it\frac {A}{n}}e^{it\frac {B}{n }})^{ n}=e^{it (A + B)}$ holds.}.
This method is the original approach taken by R.Feynman. Feynman also calculated the harmonic oscillator type-path integral, which consists of the sum of the free Hamiltonian and the external potential. In this type Hamiltonian we can calculate integration on the momentum variables, so the measure of integration would only depend on position operator．This idea is quite simple. We first divide the parameter $n$ equally [see (3.3)]. And we then multiply  same type integral kernel operator $n$ times. Then, we make  $n$ to infinity. Of course, now we make infinity operation on $n$ formally. When we exchange integral operator and the limit of $n$（ here, ``limit" means strong convergence）,``Measures of variables related to the infinite dimensional space" should be defined\footnote {For Schr\"{o}dinger type path integral, a fundamental solution of real time does not exist }.

\begin{eqnarray}
e^{-itH_{0}}{\psi}(x)&=&s-\mathrm{lim}_{n\to\infty}\Bigr(\mathrm{exp}[\frac{it}{n}H_{0}]\Bigr)^{n} \nonumber \\
&=&\Bigr(\frac{m}{2{\pi}t/n}\Bigr)^{\frac{3n}{2}}\int_{{\mathbf{R}}}\ldots\int_{{\mathbf{R}}}\mathrm{exp}\Bigr[{\frac{im}{2}}\sum^{n}_{j=1}\frac{|{x_j}-{x_{j-1}}|^2}{t/n}\Bigr]\psi(x_0)d{x_0}{\ldots}d{x_{n-1}} \nonumber \\
\end{eqnarray}

Now, when $n$ goes to infinity, we regard the part of the sum contained in the integral convergence as converging on an action functional in the sense of classical physics. This means that $S_{n} (x_0, \ldots, x_{n-1}, x, t) {\simeq}\int_{0}^{t}\mathcal {L} (x (s), \dot {x} (s))ds$. Then, as already mentioned, path integral measure becomes infinite integration.  In this case, the Lagrangian-density $\mathcal {L} $ will be the integrand.

(3.3) and (3.4) are both functions that describe the state of evolution. So, they only give the same information on the state.  And  since that integration is possible for each individual variable, the integration of (3.4) is  now formally  the same result as (3.3).

\section{The problem on time representation}
In physics, often ,we use calculation on the representation of path integrals or the integral kernel. In particular, such as String theory, we make use of these calculations . In this sense, these  calculations would be familiar.
By the way, In the representation of $e^{-itH_ {0}} $ or, $ {i}\frac{d\psi (t, x)}{dt}=H_0\psi (t, x)$, \hspace{2mm}$\psi (t_0, x)=\psi (x)$ , the time parameter $t$ is represented. what does time parameter $t$ means exactly ? Before considering that, we little need to wander.

In quantum mechanics, we know well that there is ``The problems of observation " which we called. The problems of observation is usually the issue concerning on that ,when quantum state (in this paper the system is the single particle system) observed by telescope-macroscopic , the system in non-deterministic (non-causal) will be contraction (reduction) --we will regard the change of the system as being reduction. Now I use expression ``regard", because there are some views which this reduction is non-contradiction\footnote{For example, the Copenhagen interpretation as described later  (in particular Bohr's interpretation), what the only physically meaningful is the observed phenomena, so the change in physical state earlier are not considered .}.
If we place an observer in the middle, using the terminology of Aristotle, we name the state of the quantum system before the observation ``Potential"or ``Dynamis, "and, after observing the state  ``Enerugeia"\footnote { Of course, this terminology is different from Aristotle's exact nomenclature. }. The observation problem is the issue that  how we consider the relationship between Dynamis (from now we name it A or A-world ) and Enerugeia (we name it B or B-world) througt  ``Observer" (we name it C). 
The interpretation of quantum physics which is commonly understood today is  ``Copenhagen interpretation " by N.Bohr and W.Heisenberg , et al. Also on the Copenhagen interpretation, there is a variety of historical change. Core concepts of the Copenhagen interpretation is, due to Heisenberg, ``Uncertainty relation "\footnote{【Uncertainty relation】：

Suppose $T$,$S$ are symmetric operators on$\mathcal{H}$, and $\psi{\in}D([T,S])$,
then

\hspace{35mm}$(\Delta{T})_{\psi}(\Delta{S})_{\psi}{\ge}{\frac{1}{2}}\mid(\psi,[T,S]\psi){\mid}$

Remark;  $(\Delta{T})_{\psi}$is Standard deviation of radom variable $T_{\psi}$.}. 
Also, by Bohr ``Complementarity "\footnote { In a lecure , 16, September, 1927, Como (Italian city) , Bohr first went public about the idea of complementarity. At this time, Bohr said, ``space-time and the requirements of causality is complementary to each other .} is important. The general important points of the Copenhagen interpretation are the following two points.

(α）Phenomenon of quantum physics and the observer is inseparable\footnote { In Bohr's interpretation, the observer interacts with the state of the system. thus ,the observations before which is the states contracts eigen-stateis meaningless. }.

(β）Phenomenon of quantum physics is told in the language of classical physics. Therefore, the picture of wave and particle is impossible in the category of classical physics. But in fact, several (dual) pictures of their imaging helping each other, we will be able to talk in the language of classical physics quantum physics phenomenon.

Now, ``A, B, C" distinction was made. This distinction does not come from Bohr himself. It is said that rather Heisenberg incorporated the idea that a quantum system will change from Dynamis to Enerugeia into the concept of complementarity\footnote{ For Bohr, complementarity did not hold between the possibility and impossibility to observe. when we specify all conditions of observation, complementarity is meaningful according to Bohr. Heisenberg maybe had a different meaning to use Bohr's complementarity .}.
Bohr and Heisenberg are different. It seems Heisenberg  admitted to the mathematical meaning to describe the function of previously observed states\footnote{ I found statements attest to this point in the work of Heisenberg ( [7], S.330-333). It seems that he thought the extension of ordinary logic in the mathematical principle of complementarity and mathematical ontology. In addition, he cited Plato's \textit{Timaeus} and he said that elementary particles are represented by mathematical form. By contrast, Bohr had acknowledged the importance of mathematics, but, speaking in relation to quantum physics, he said that mathematics was only a logical framework in the philosophy of nature and abstract form ( \textit{mathematics and natural philosophy} (in his Lecture 1954). }.

The current mathematical physics considering the strong relationship between the mathematical theory and theoretical physics, Heisenberg's interpretation will be useful when discussing time. And together with the theory and Bohr's complementarity and Heisenberg's theory of complementarity, currently has been built the modern Copenhagen interpretation.

Now, the problem here is the following. Which world (phase) does the parameter $t$ in quantum theory or Beginning of space-time belong to, in the case of according to Heisenberg's theory of measurement? When we suppose that A-world is being in different sense from B-world, in the case that A does not logically led to the collapse for B, there is no problem.

However, if we assume that when macroscopic space-time world that we are feeling  emerges through the observation ,  From ``mathematical and metaphysical" view point, the meaning of the parameter $t$ is room to reconsider for us.

As Heisenberg said, we suppose to exist the quantum A-world. In this case, can we interpret the time variable parameter of the state functions on Hilbert space (and, as discussed a little later, including the spatial coordinates), as the time variable parameter  of classical physics,  ``naively" ?
Here ``naively" means to, whether the time parameter may be understood as the real elements (or, elements of a real function ) .

\medskip
I think about how the relationship between quantum physics and time. These days it seems that basic research is increasing on the relations. Among them, the most notable is the view that the time is thought of as operators.

\section{Time operator}

If time is simply a scalar quantity, because a Hamiltonian is defined on the Hilbert space in general (unbounded) operator, so the canonical commutation relation (CCR)  becomes $[t, H] = 0$. However, as a consequence of the experimental evidence that in the unsteady state energy measuring we have the variation in measurements, we may understand the uncertainty relation between energy of the system and the time on the width of the steady-state [9].
But time is, as just mentioned, usually, judging from the analogy of classical physics, a scalar quantity mathematically, and we should not see it as the operator. Therefore, the operator on time have been considered non-existent. But recently, the idea that quantum time should be reformulated as operator.

If we consider time as operator, at just the Hamiltonian of the Schr\"{o}dinger type, it is not self-adjoint operator. This fact is easy to see from ``von Neumann uniqueness theorem" that there is no self-adjoint operator satisfies CCR with the self-adjoint operator bounded below [10].
Therefore, if time operator were self-adjoint, of course, according to von Neumann uniqueness theorem, its existence is denied. Of course, now, we suppose the Hamiltonian is bounded below. Because in the real world , in our universe, if the ground state energy would not exist,  our phenomenal world would collapse.
But we do not know the idea that the Hamiltonian is bounded apply the theory of the Beginning of space-time. Rather, in the Beginning of space-time maybe there is no ground state energy.

The time on classical physics is often likened to the arrow of time. Also it is interpreted as a linear time ranging from the past to the future. However, this time picture is measured in metric scale similar to our bodies. It would be macro time in the B-world.  We have a problem with the interpretation that macro time would be to expand the quantum world.
We think classical physical time, indeed, as a one-dimensional, and  as also  flowing seamlessly. We can make the images of this time spatial. That time would also relate to our perception of moving objects (From physical view, state of standing still is moving state.) Motion of an object is , in a classical physics, at just limiting to the one-dimensional motion, which must be continuous, not discrete (not quantum).
The state of motion is uniquely determined. In addition, the momentum (or energy \footnote { The energy can be defined using the momentum.}) and time is deterministic  separately.

However, in quantum systems the situation is not so simple. The uncertainty relation indicate that position and momentum has not separately and independently the final value. Also, as mentioned,  the time and energy uncertainty physically has been formulated.

Time operator is the operator satisfies CCR with the Hamiltonian of the system.  The reason why the operator is said that the time operator is that the shape of the Aharonov-Bohm operator which is original form of the time operator formally has physically time-dimension [5].
In quantum physics, observable (operator) and the value of observations (spectrum) should be considered separately. If an observables would be not self-adjoint operator, the value on the observation would not have a real value. Therefore, because the belief that time must be real, since the Pauli claimed, has been considered the time operator does not exist. However, according to Heisenberg's uncertainty relation, the two operator of satisfied CCR together not necessarily self-adjoint operator. We can extend the CCR to symmetric operator rather than self-adjoint operator.

\medskip

Time operator is defined as the symmetric operator\footnote{ Strictly speaking,  symmetric extension done. Rigorous mathematical formulation was made by Miyamoto [16 ]. }, and, in general, time operator is defined as a symmetric operator satisfying CCR relation  with the Hamiltonian of the system .

【Definition of time operator】

On a Hilbert space, $T$ is the time operator for the Hamiltonian $H$ is satisfying the following.
\medskip

 (1） $T$ is symmetric operator.
 (2） On $D(TH){\cap}D(HT)$, $[T , H]=i$ is satisfied.

\medskip

Time operator gives information about the Hamiltonian and time evolution of the system, through analysis of CCR\footnote{ For this analysis, it is necessary to add the following definition.

\medskip

【Definition of $T$-weak Weyl relation】

\medskip

Let $T$ be symmetric operat and $H$ be self-adjoint operator on $\mathcal{H}$．Then, That $T$ and $H$ satisfies $T$-weak Weyl relation is for all $\psi\in{D(T)}$,and for all$t\in\mathbf{R}$, $e^{-itH}\psi\in{D(T)}$, and

\hspace{35mm} $Te^{-itH}\psi=e^{-itH}(T+t)\psi$

holds．

\medskip

Suppose that $T$ and $H$ satisfies "Definition of $T$-weak Weyl relation". In addition to this, $H\ge{0}$ and, For all  $t\in\mathbf{C}$($\mathrm{Im}t{\le}{0}$, $\mathbf{C}$ is Complex number．) on "Definition of $T$-weak Weyl relation, the uncertainty relation on Time and Hamiltonian is as following．

\hspace{35mm}$(\Delta{T})_{\psi}(\Delta{H})_{\psi}>\frac{1}{2},  \hspace{5mm} \psi\in{D(TH){\cap}D(HT)}, \hspace{3mm}(\|\psi\|=1)$

it is important that equality is not satisfied．

\medskip

For Hamiltonian $H=P^2$, Aharonov-Bohm time operator of $T$ has the form defined as follows．

\hspace{35mm} $T:=\frac{1}{2}(QP^{-1}+P^{-1}Q)$,

\hspace{5mm}(domain is $D(T):=D(QP^{-1}){\cap}D(P^{-1}Q)$ )

Therefore, for $H_{0}$, Time operator $T_{0}$is

\hspace{35mm}$T_{0}:=\frac{m}{2}(QP^{-1}+P^{-1}Q)$.

By expanding this Symmetric operator, we can composed of time operators satisfies $T$ -weak Weyl relation.}.

\bigskip

Currently, through the basis for the representation theory of unbounded operators of K.Schm\"{u}dgen [14], a generalization of time operators has done by A.Arai et al. Prior to that, contrary to the objection of Pauli, Galapon, under additional conditions, found a self-adjoint time operator with real spectrum [2][6]. Therefore, the discussion on time operator  has shown new developments.

According to Arai, Aharonov-Bohm type's time operator is the essential self-adjoint operator ,in which many operators are classified by the weak Weyl relation in some class (of closed strong means time operators)\footnote{ If the Hamiltonian $H$ is bounded below or above, Time operator $T$ accompany to $H$ is not the essentially self-adjoint is known. Therefore, if the Hamiltonian is unbounded, time operator can be a self-adjoint operator. However, in the case that the Hamiltonian is unbounded, that time operator theory might not  be helpful to understand the reality of our universe. However, it is mathematically possible.}. In addition, the some time operator type which belong to  wider class  known to us\footnote{ The classification of time operator depend on the domain which time operator and Hamiltonian belong to  and on the topology class.}.

By the way, the spectrum of time operators, in the relation to the Hamiltonian $H$ accompany time operator, can be  stochastically calculated as an average of $T$. Using the inner product of the Hilbert space ; that is, $(\psi, T\psi): = <T> = t$.
When the time operator is not essentially self-adjoint ( ie, If it would be symmetric operator), the spectrum of time operator  (in this case which the spectrum is discrete, it also called eigenvalues) is complex number [12][13].  Also, if the time operator is Galapon type, the spectrum of time operator is real under some conditions.
When we organize various type time operators,  even if  we include such a case that the Hamiltonian with time operator can not appear directly in realistic physical value, mathematically, the time operator can be  observable operator.

Here we would like to think about ``the two worlds of Heisenberg", which means ``the structure concerning on the A-world and B-world across the observer C .

Time operator of the A-world  may be naturally different from the scalar quantity. In other words, through taking the average value of operators and computing the standard deviation between the state vector (looking upon  this  working as C), time may appear  as scalar ``$t$". By the uncertainty relation, it will means that in A-world time is only talk as a kind of  ``fluctuation". Here, the image of classical physical time might be bankrupt．In A-world, we can only describe time as the whole neighborhood on some time, but not pointing it correctly. In other words, we could only express time as the stochastic event.

\section{A perspective on quantum time representation and time operator }

As shown in Section 3, we reconsider what the implication might be parameter $t$, $x$ of the equations governing the behavior of quantum particles states in Hilbert space.

\hspace{35mm}${i}\frac{d\psi(t ,x)}{dt}=H_0\psi(t, x)$  ,\hspace{5mm}  $\psi(t_0, x)=\psi(x)$

Same form as this equation parameters is also used,

 $\psi (t, x) = U (t, t_0)\psi (t_0, x) $, and on $U (t, t_0) = \mathrm {exp} (- {i} H_0 (t, -t_0)) $ , these parameters has dictated the content of path integral representation or of the integral kernel representation.

Let B be physical macroscopic classical world, and  A be  quantum world. We usually  suppose the parameter $t$ or  $x$, regarding as the same kind common parameters in  A and B and  calculate. This will be correct?  For a particle behavior in the A-world, why we can use the time parameter is in macro world B. Needless to say, the reason why, if  A exists (according to Bohr, the ontological means of A-world is nonsense), we actually observe the physical data in B- world. If we stand in a narrow empirical position, the existence of A-world is hypothesis forever, according to``making any unnecessary hypothesis" of principle of W.Ockham (1285-1349) or I.Newton (1642-1727). If A-world itself exist or be regarded as merely a logical construct, be regarded as fictional reality in the physical values  coming out through the observation, there is not differences view\footnote{ Speaking from this sense, we can say that mathematicians and physicists would catch the world of complex number or  Hilbert spaces, in different view point. In this sense, A world might be a mathematical world }.

By the way, but in this paper we suppose the two worlds theory of Heisenberg, more broadly, the two world theory since Plato, so we would like to discuss on accepting theoretical possibility of the existence of the A-world\footnote{ However, because the Platonic two worlds theory is built by non space-time relationship with the space-time world, so it is different from Heisenberg's two world theory slightly. }.

\medskip

We suppose A is quantum world and this world obey to rule of quantum world's parameter ,not obey to the rule of the classical parameter. If A-world not be shown up us  only through  C, we would only be able to talk about time parameter $t$ according to stochastic process.

Formally, if modified, then the equation above,,

\hspace{35mm}${i}\frac{d\psi(<T>, <X>)}{d<T>}=H_0\psi(<T>, <X>)$, \hspace{5mm}  $\psi(<T_0>, <X>)=\psi(<X>)$
then,

$\psi(<T>, <X>)=U(<T>, <T>_0)\psi(<T_0>, x)$, $U(<T>, <T_0>)=\mathrm{exp}(-{i}H_0(<T> ,-<T_0>))$. Here $<T>$, $<X>$ represents the average time, position of the operator.

Once that happened reduction of wave packet, one parameter is determined, when the parameter become a real number, Can we the number regard  as the classical parameter governing the classical B-world ?
 
 However, that the average time operator (spectrum) will be real is  a very special case. As far as the Aharonov-Bohm type, we  need removing the boundedness condition of that Hamiltonian.

 \bigskip
Therefore, the following, we will consider the three cases ([1][2][3][4][6]).

\medskip
(1) Aharonov-Bohm time operator (an example of a (closed) strong time operator);

This type, as already mentioned, is the earliest form appeared as a time operator. Pauli pointed out that, when the Hamiltonian is unbounded, the operator will not become a self-adjoint, so the spectrum of the operator is not real but become complex.
 
Let me assume the following;  A-world might be unbounded Hamiltonian world. Rather, A-world might be a world of the Beginning of space-time beyond logic of the present space. If so, on fluctuations in some indeterminate state the world might exist that we could not determine the  bound of energy. It may be almost ``Black-hole" world. However, through the relationship between the observer the A-world appears and the phenomenal B-world at the moment, a Hamiltonian might be bounded. At this time, the spectrum of time operators will become complex. If a result of the A-world nature on the Hamiltonian might  propagate to the result of B-world,  a part of the spectrum would become a real parameter .
Tim operator is also ,if it is weak Weyl-type, is essentially self-adjoint. In this case, by von Neumann's uniqueness theorem, the representation of the Hamiltonian and time operator will be equivalence of the Schr\"{o}dinger representation. So the relationship between the time operator and the Hamiltonian is the same as the relationship between momentum and position.
This view is close to the interpretation of Aristotle, Leibniz. In other words, the interpretation on time is closer to the view that  ``time can be the first there when  movement exist".

\medskip

(2) Galapon's time operator (as an example of harmonic oscillator type);

In the case of Hamiltonian is, $H={\frac{1}{2m}}P^{2}+{\frac{1}{2}m{\omega}^2{x}^2}$,($\omega$; Angular frequency), the spectrum of time operator $\sigma(T)$ is in interval $[-\pi/\omega,  \pi/\omega]$, therefore, is real number. Now, momentum of the Hamiltonian and position operator are equivalent to the Schr\"{o}dinger representation. What does the Spectrum in real time and its to be limited to the interval mean?

\medskip　

(3) In the case of the spectrum is complex (time operator is not essentially self-adjoint  operator );　
\bigskip

As mentioned in (1), time operator with spectrum of the complex number has meaningful ``Wiener measure", when it is represented as a path integral. In functional integral theory, which is connected to the discussion on the heat kernel semigroup.

\begin{equation}
e^{-tH_{0}}{\psi}(x)={\textrm{lim}}_{R{\to}\infty}\Bigr(\frac{m}{2{\pi}|t|}\Bigr)^{\frac{3}{2}}{\int}_{|y|{\leq}{R}}e^{-{\frac{m}{2t}}|x-y|^{2}}\psi(y)dy
\end{equation}

In the case $t>0$, (6.1) is bounded. With this integral kernel, the path integral, by the Feynman-Kac formula, can change the representation of stochastic process. Of course, in this case, the parameters remain in classical representation. But, at least, imaginary-time and real-time, on path integral representation, has raised mathematically important issue related to the existence of a measure of integration\footnote { This issue is also related to ``the imaginary time theory" of S. Hawking .}.

\section{Application of four-dimensional relativistic space-time representation}

On Minkowski space-time which is a 4-dimensional space-time of special relativity, the world length $s$ is determined by indefinite metric inner product [8]. Nevertheless, the parameters appearing in $s^2=- (ct)^2+x^2+y^2+z^2$ is in B-world from the quantum space-time view of the A-world. Therefore, we may also need to think of the structure $ <T> + <X> + <Y> + <Z> $ .
However, the differentiable Riemann space-time is assumed to be smooth. But if we put a quantum structure, as is well known, there should take a non-commutative structure of space-time. The curved space theory and the Gaussian surface theory does not exist in quantum theory. Quantization of general relativity is that, mathematically, still close to the most transcendental problems.
Even the problem of time operator on regular metric space, flat space, yet there are many problems. That we extend the theory of time operator to the indefinite metric space (such as Minkowski space) and  also  to a curved space, will be required  hard work.

About the famous ``gravity equation"

\begin{equation}
   {R_{\mu \nu}}-{\frac{1}{2}}{g_{\mu \nu}}R-{\lambda}{g_{\mu \nu}}={\eta}{T_{\mu \nu}} 
 \end{equation},
 
 \medskip
 
 ($ \eta $ is a constant related to the gravitational constant G, $\lambda $ is Einstein's gravitational constant to discuss the structure of the universe, ${R_ {\mu\nu}} $ is the quantity made from the curvature tensor, $R$ is called the curvature scalar quantity, $T_ {\mu \nu} $, the energy-momentum tensor.).
 
 \medskip
we will need to generalize the quantity on Riemannian metric tensor $g_{\mu \nu}$.

As long as they are the space-time measure, its structure also may be required to the generalized measure beyond space and time. Riemannian metric is only to determine the geometric properties of space, so it is, from the standpoint of classical physics, equivalent to the gravitational potential of Newtonian mechanics. Also by the structure of potential in the extreme world,  The representation of metric maybe should be reconsidered. Of course, attempts to canonical quantization of gravity has been made [17], the theory may need more than the current metric form.

Moreover, quantum theory is essentially on the structure of linear space. As like renormalization problem, we will have to consider the problem of quantum chaos in a nonlinear structure (such as the gravitational field equations).

 Besides Platonic view, such as also found in Japanese mythology, the universe Genesis is essentially the world might be a transition to space-time cosmos from the chaotic world .
\section{Conclusion }

As a theory on physics of our universe, needless to say, the relativistic theory must be important and for Genesis theory on our micro universe the quantum theory is also necessary.
In the beginning of the universe, four powers had been unified under the higher symmetry. Superstring theory has been discovered from the aspects of the unified theory , as well. The net spin theory is found from the quantum gravity. In the theory of spin net ,the existence of space-time does not be  assumed [17]. As long as the spin network theory, unlike superstring theory, the parameters on space-time do not  embedded into the vibration of strings and membranes.

By the way, as the Genesis  modern story of our universe, it is said that the Big Bang universe had  hold low-entropy state with higher symmetry, so to speak, hold to be a beautiful original condition. But the Beginning of space-time was relally smooth ?
If the Beginning of space-time was a world without space-time, from where space-time was born? As ``imaginary time cosmology by Hawking", if the origin of space-time had no beginning and the imaginary-time (time of imaginary number) in the first, imaginary-time would have been transformed into real-time and the space-time universe would present ?
 
Complex number , as is well known, has no meaning, when considering the distance in our world. When complex number  is taken the square root of the product of the conjugate, we can measure something. If the primordial world would have been complex, the beginning of space-time might be in complex world beyond this space-time. The primordial world may be regarded as the world that previous world before observation in the region of space-time.
The current age of the universe around 13.7 billion years, the curvature is much flatter, Most of the material constituting universe is the dark matter (22 percent), dark energy (74 percent), and the baryonic matter is 4 percent. As dark energy and dark matter, neutrinos with mass have been cited with. Dark energy may have a logical structure of mathematical supersymmetry beyond the fermions and bosons.

On the issue of quantization of space-time, with the probabilistic interpretation, under the binding conditions, we will also encounter the issue of space-time discreteness. What is discrete space-time ? Why do we have 4-dimenton? [15].  If our bodies (including brains) to be regarded as continuous monitoring equipment for the outside world (A-world), that the brain has a digital phenomenon is  perhaps due to the discrete time and space in the brain.

A world of absolute continuity in mathematical scattering theory corresponds to the continuous spectrum of the free Hamiltonian. The Beginning of space-time might have  been continuous and ,then our present universe had been born, the observation instruments  (our selves)  was born, so ,in fact, in this world  space-time may be discrete . We may interpret space-time as to be continuous, but in fact, this view may be only  depending on the ability of our sense.

\section*{Acknowledgment}

This work was supported
by members of the Research group on the Beginning of the space-time in Hokkaido University.

\end{document}